\documentclass[12pt]{article}
\usepackage{amssymb}
\usepackage{graphicx}
\usepackage{amsmath}

\def\d{\partial}
\def\l{\left(}
\def\r{\right)}

\newcommand{\be}{\begin{equation}}
\newcommand{\ee}{\end{equation}}
\newcommand{\ba}{\begin{align}}
\newcommand{\ea}{\end{align}}
\newcommand{\bg}{\begin{gather}}
\newcommand{\eg}{\end{gather}}
\newcommand{\bseq}{\begin{subequations}}
\newcommand{\eseq}{\end{subequations}}

\renewcommand{\exp}{\mathop{\rm exp}\nolimits}

\def\half{\frac{1}{2}}

\begin{document}

\title{Are $R^2$- and Higgs-inflations really unlikely?}

\author{
D.~S.~Gorbunov$^{1,2}$\thanks{{\bf e-mail}: gorby@ms2.inr.ac.ru}\; and 
A.~G.~Panin$^{1}$\thanks{{\bf e-mail}: panin@ms2.inr.ac.ru}\\
\mbox{}$^{1}${\small\em Institute for Nuclear Research of Russian Academy of
  Sciences, 117312 Moscow,
  Russia}\\  
\mbox{}$^{2}${\small\em Moscow Institute of Physics and Technology, 
141700 Dolgoprudny, Russia}
}
\date{}

\maketitle

\begin{abstract}
We address the question of unlikeness of $R^2$- and Higgs inflations
exhibiting exponentially flat potentials and hence apparently
violating the inherent in a chaotic inflation initial condition when
kinetic, gradient and potential terms are all of order one in Planck
units. Placing the initial conditions in the Jourdan frame we find
both models not worse than any other models with unbounded from above
potentials: the terms in the Einstein frame are all of the same order,  
though appropriately smaller.   
\end{abstract}


The $R^2$- and Higgs-inflation models~\cite{starobinsky,Bezrukov:2007ep} 
work in a very economic way.
Inflationary stage is attained due to modification of the
gravitational sector with respect to the presently accepted 
paradigm---General Relativity and Standard Model of particle 
physics---by introducing a quadratic curvature term in $R^2$-inflation 
and a strong nonminimal interaction with the Higgs boson in
Higgs-inflation. In the Einstein frame, where the gravity
takes the Einstein--Hilbert form, this stage is realized as 
a slow-roll inflation at super-Planckian values of a
corresponding scalar
degree of freedom in each theory. The scalar has very flat 
plateau-like potential at large values of the field bounded from above 
by the value of 
\begin{equation}
\label{V-at-inflation}
V_0 \simeq 10^{-12}\times M_{\text{Pl}}^4\,,
\end{equation} 
which comes from fitting to the matter power spectrum extracted from
the CMB anisotropy map \cite{Bennett:2012zja,Ade:2013nlj} and 
observations of the Large Scale Structure  
\cite{Jones:2009yz,Abazajian:2008wr}. 

In this work we address the issues of formulated 
in Ref.~\cite{Ijjas:2013vea} ``unlikeness'' problem  
to the $R^2$-inflation and Higgs-inflation. 
This problem originates in the initial conditions for a successful 
inflation. According to
Refs.~\cite{Linde:1983gd, Linde:1984ir} for a chaotic inflation 
to begin in a relatively uniform domain of the Planckian size it is 
sufficient to have {\em each} of the forms of inflaton energy density being
of the Planckian value, 
i.e.\footnote{To simplify the formulas hereinafter, we omit the scale
  factor; it can be straightforwardly recovered depending on the
  choice of metric.} 
\begin{equation}
\label{inf-initial}
\frac12\dot{\phi}^2 \sim \frac12(\d_i \phi)^2 \sim
V(\phi) \sim M_{\text{Pl}}^4\,.
\end{equation}
However, for the models with plateau-like
potential, $V(\phi)\simeq V_0$, the initial condition one might  
expect is   
\[
\frac12\dot{\phi}^2 \sim \frac12(\d_i \phi)^2 \sim M_{\text{Pl}}^4 \gg
V(\phi)\,.
\]
In this case the kinetic and gradient inflaton energy densities  
quickly dominate and hinder 
the inflation from launching. The Universe 
starting with these initial conditions undergoes inflationary
expansion later, only if the initial uniform space region  
would be much bigger than the Planckian domain. Such initial homogeneity of the
Universe looks very unnatural and ``unlikely''.

We start consideration of this problem 
 (see e.g. \cite{Starobinsky:1982mr,Starobinsky:1987zz} 
for earlier studies)  
with $R^2$-inflation. 
Lagrangian of this theory in the Jordan frame (JF) 
is\,\footnote{Following \cite{Gorbunov:2011zz} 
we choose the Landau--Lifshitz convention for 
the variables in the gravity sector, so that 
$(+,-,-,-)$ is the metric signature, and the scalar curvature is
negative at inflationary stage.} 
\be
\label{SJF}
S^{JF}=-\frac{M^2_{\text{Pl}}}{16\,\pi}\int \!\! \sqrt{-g^{JF}}\,d^4x\, 
\left [ R^{JF}-\frac{(R^{JF})^2}{6\mu^2} \right ]\;.
\ee
It is convenient to go to the Einstein frame (EF) where the 
gravity action takes the Einstein-Hilbert form and the scalar degree 
of freedom responsible for the inflationary solution becomes canonically 
normalized. For this purpose we introduce a new auxiliary scalar
field $Q$ and write the action~\eqref{SJF} in the following way 
\cite{Hindawi:1995cu}
\be
\label{SJFQ}
S^{JF}=-\frac{M^2_{\text{Pl}}}{16\,\pi}\int \!\! \sqrt{-g^{JF}}\,d^4x\, 
\left [
\l 1-\frac{Q}{3\mu^2}\r \l R^{JF} - Q\r +
\l Q-\frac{Q^2}{6\mu^2}\r
\right ]\;.
\ee
Varying \eqref{SJFQ}  with respect to $Q$ we obtain 
\be
\label{Q}
Q = R^{JF}\;,
\ee
and eq.~\eqref{SJFQ} reduces to original action~\eqref{SJF}. 
Then we get rid of the factor 
\be
\label{factor}
\Omega^2=1-\frac{Q}{3\mu^2}
\ee
by proper rescaling, that is by making conformal transformation of the metric  
\be
\label{EFmetric}
g_{\mu\nu}^{EF} = \Omega^2\,g_{\mu\nu}^{JF}\;, \quad \text{with}\quad
\Omega^2 \equiv \exp\l\sqrt{\frac{\pi}{3}} \frac{4\,\phi}{M_{\text{Pl}}}\r\;.
\ee
Thus we arrive at the following action in the EF
\be
\label{SEF}
S^{EF}=\int \!\!\sqrt{-g^{EF}}\,d^4x\,\left[
  -\frac{M_{\text{Pl}}^2}{16\,\pi}
\,R^{EF} + \half\,
g^{EF}_{\mu \nu} \d^\mu \phi \d^\nu\phi
-\frac{3\,\mu^2\,M_{\text{Pl}}^2}{32\,\pi}\, \l 1-\frac{1}{\Omega^2} \r^2
\right]\;.
\ee

The homogeneous and 
isotropic Universe described by the action~\eqref{SJF} undergoes
an inflationary expansion at large values of $R^{JF}$. In the EF  
this stage is realized as slow-roll inflation taking place at
large values of the scalar field, $\phi\gg M_{\text{Pl}}$ which 
plays the role of the
inflaton. Normalization of the amplitude of the generated during
inflation scalar perturbations 
 to the spectra of observed CMB anisotropy 
and Large Scale Structure \eqref{V-at-inflation} 
yields the estimate $\mu \approx 2.5\times 
10^{-6}\times M_{\text{Pl}}$.  

Now let us discuss initial conditions for this model. Looking at the
action in the EF~\eqref{SEF} one may conclude that the theory suffers
from "unlikeness" problem raised in Ref.~\cite{Ijjas:2013vea}. Indeed,
at large $\phi$ the potential is very flat and tends to the constant
\eqref{V-at-inflation}, $V\to V_0 = 3\mu^2M_{\text{Pl}}^2/(32\,\pi)
\simeq 10^{-12} \times M_{\text{Pl}}^4$.  To clarify this question let
us formulate a natural initial condition in the JF which is the
physical frame of the theory. In this frame the model is described by
pure gravitational action \eqref{SJF}, thus in the sense of
Refs.~\cite{Linde:1983gd, Linde:1984ir} one expects that the Universe
in a Planck scale domain has the Planckian curvature
\begin{equation}
\label{Jordan-curvature}
|R^{JF}| \sim M_{\text{Pl}}^2\;. 
\end{equation}
According to eqs.~\eqref{Q}, \eqref{factor}, for
the transformation function $\Omega$ connecting the two frames we have
\begin{equation}
\label{Q-value}
\Omega \sim M_{\text{Pl}}/\mu \sim 10^{6}\;. 
\end{equation}
This leads to $\phi \sim
|\log(M_{\text{Pl}}^2/\mu^2)| \sim 20$ for the initial inflaton field
value and $V(\phi) = V_0 \simeq 10^{-12} \times M_{\text{Pl}}^4$ for
the initial potential energy, as required, see \eqref{V-at-inflation}.

In order to estimate the initial kinetic and 
gradient energy of the inflaton in the Einstein frame 
one may use the relation
\be
\label{R-R}
-R^{JF} = -\Omega^2\,R^{EF} + \Omega^2\,\frac{8 \pi}{M_{\text{Pl}}^2}
g_{\mu\nu}^{EF}\d^\mu \phi \d^\nu \phi - \Omega^2\,\frac{4\sqrt{3
    \pi}}{M_{\text{Pl}}} 
\,g_{\mu\nu}^{EF}\d^\mu \d^\nu \phi \;,
\ee
between Ricci scalars in the two frames. The first two terms in
r.h.s. of eq.\,\eqref{R-R} refer to the pure gravity and scalaron, 
whose initial contributions to the the energy density are
of the same order, according to Refs.~\cite{Linde:1983gd,
  Linde:1984ir} and equations of motion in the EF. Hence 
\[
\left | R^{JF} \right |\,
\Omega^{-2}\sim \left | R^{EF} \right | \sim \dot\phi^2\,M_{\text{Pl}}^{-2} \sim 
(\d_i\phi)^2\,M_{\text{Pl}}^{-2} \,.  
\]
Recalling \eqref{Jordan-curvature} we get 
$\frac12\dot{\phi}^2 \sim \frac12(\d_i \phi)^2 \sim M_{\text{Pl}}^4\,
\Omega^{-2}$. Then taking the potential $V(\phi)$ from \eqref{SEF} and
inserting the estimate \eqref{Q-value} we obtain finally 
\[
  \frac12\dot{\phi}^2 \sim \frac12(\d_i \phi)^2 \sim V(\phi) 
\sim \mu^2 M_{\text{Pl}}^2/30\sim 10^{-12} \times M_{\text{Pl}}^4\,.
\] 
But it is just what is needed for a successfully inflationary model,  
see eq.\,\eqref{V-at-inflation}. We observe, that in the
Einstein frame all the relevant terms start with the same initial
value, which is however smaller, than the Planck mass.

One can find the very similar
arguments working well for the Higgs-inflation, which is not worse
than $R^2$-inflation in this respect.  
The action of the Higgs-inflation model in the JF is~\cite{Bezrukov:2007ep} 
\be 
\label{Higgs-JF}
S^{JF}=\int \!\! \sqrt{-g^{JF}}\,d^4x\, 
\left [ -\frac{M^2_{\text{Pl}}}{16\,\pi} 
\l 1 + \frac{8\,\pi \xi h^2}{M^2_{\text{Pl}}}\r R^{JF} + 
\frac12 g^{JF}_{\mu \nu}\d^\mu h \d^\nu h - \frac{\lambda}{4}h^4 
\right ]\;,
\ee
where we use the unitary gauge with $h$ being the Higgs boson. 
Going to the EF by conformal transformation~\eqref{EFmetric} with
\be
\label{Higgs-omega}
\Omega^2 = 1 + \frac{8\,\pi \xi h^2}{M^2_{\text{Pl}}}
\ee 
we arrive at the following action in the EF
\be
\label{Higgs-EF}
S^{EF}=\int \!\!\sqrt{-g^{EF}}\,d^4x\,\left[
  -\frac{M_{\text{Pl}}^2}{16\,\pi}
\,R^{EF} + \half\,
g^{EF}_{\mu \nu} \d^\mu \phi \d^\nu\phi
-\frac{\lambda}{4}\, \frac{h(\phi)^4}{\Omega(\phi)^4} \right]\;,
\ee
where we replace $h$ with canonically normalized 
scalar field $\phi$ utilizing the relation 
\be
\label{h-phi}
\frac{d \phi}{d h} = \sqrt{\frac{\Omega^2 + 
48\pi \xi^2h^2/M_{\text{Pl}}^2}{\Omega^4}}\;.
\ee
Inflation in this model in the JF happens at large values of 
$h \gg M_{\text{Pl}}/8\pi \sqrt{ \xi }$. In this limit from
eq.~\eqref{h-phi} we get 
\be
h \simeq \frac{M_{\text{Pl}}}{\sqrt{8 \pi \xi}}
\exp \l 2\sqrt{\frac{\pi}{3}}\, \frac{\phi}{M_{\text{Pl}}}\r\;.
\ee  
In this case the inflaton potential in the EF becomes exponentially flat
and takes form (cf. eq.~\eqref{SEF})
\be
\label{Higgs-pot}
\frac{\lambda}{4}\, \frac{h(\phi)^4}{\Omega(\phi)^4} \simeq
\frac{\lambda M_{\text{Pl}}^4}{256\, \pi^2 \xi^2}
\left [ 1+ \exp \l - 4\sqrt{\frac{\pi}{3}}\, \frac{\phi}{M_{\text{Pl}}}\r
\right ]^2\;.
\ee   
Normalization of the amplitude of the scalar perturbations generated in
this model during inflation yields the estimate 
\begin{equation}
\label{mixing}
\xi \simeq 47000 \sqrt{\lambda} \sim 1.5 \times 10^4\;.
\end{equation} 
This implies 
\be
\label{need-2}
V_0 \simeq 1.8
\times 10^{-13} \times M_{\text{Pl}}^4
\ee
for the upper bound.     

We start the discussion with initial conditions in the JF
\eqref{Higgs-JF}. In the sense of Refs.~\cite{Linde:1983gd,
  Linde:1984ir} (see Ref.\,\cite{Barvinsky:1994hx} for an alternative
approach) one may expect that the Universe starts with near
Planckian values of the energy density in all the species. For the
potential energy we immediately get 
\begin{equation}
\label{higgs-energy}
\lambda h^4/4 \sim M_{\text{Pl}}^4\,,
\end{equation} 
and hence $h \sim M_{\text{Pl}}$. It
corresponds to the plateau-like part of the scalar field potential in
the EF, $V(\phi) \sim V_0 \simeq 10^{-13} \times M_{\text{Pl}}^4$ and
$\Omega^2 \sim 5 \times 10^6$.  But with the kinetic energy density
the situation is not so clear due to nonminimal coupling to
gravity in \eqref{Higgs-JF}. 
As soon as Ricci scalar $R^{JF}$ contains derivatives, this
term, proportional to $\xi$ \eqref{mixing},  
represents large mixing between kinetic term of the scalar field
and metric derivatives. One can make it explicit, for example, by
expanding the curvature above the Minkowski background. 
From eqs.~\eqref{h-phi} and \eqref{mixing} one
observes that this term gives the main contribution to the kinetic part
of the action for $\phi$ in the EF. 

Moreover, a non-zero value of $h$ rescales the gravity mass scale: 
$\Omega M_{\text{Pl}}\to M_{\text{Pl}}$, see \eqref{Higgs-JF}, 
\eqref{Higgs-omega}. Then, following Refs.~\cite{Linde:1983gd,
  Linde:1984ir}, the gravity contribution to the 
total energy density, the scalar kinetic terms and the scalar 
potential are expected to be of the same order, which yields the 
reliable initial conditions
\begin{equation}
\label{JF-conditions}
\Omega^2\,M_{\text{Pl}}^2\,R^{JF}\sim \xi \dot h^2 \sim \xi 
 (\d_ih)^2  \sim\lambda h^4 \;.
\end{equation}
Adopting the estimate \eqref{higgs-energy} we obtain then 
\begin{equation}
\label{JF-curvature}
R^{JF} \sim M_{\text{Pl}}^2/\Omega^2
\end{equation}
consistently with the equation of motions in the JF.   
Using the relation
between the Ricci scalars in the two frames, 
which in the case of large $\Omega$ 
coincides with eq.~\eqref{R-R}, we arrive at the following 
initial conditions in the EF
\[
\frac12 \dot{\phi}^2 \sim 
\frac12(\d_i \phi)^2 \sim M_{\text{Pl}}^4/\Omega^4 \sim 10^{-13}
M^4_{\text{Pl}}\,, 
\]
that is right what we need, eq.\,\eqref{need-2}. 

We emphasize that in
each frame all the terms, when correctly identified, are of the same
order, but the scales in the JF and in the EF are different, which
reminds the situation in $R^2$-inflation. In the Higgs-inflation,
since at large $\Omega$ the value of $\sqrt{8\,\pi\xi}$ replaces the
Planck scale in \eqref{Higgs-JF}, one can introduce any value
$\Lambda$ to be utilized instead of $M_{\text{Pl}}$ in
eq.\,\eqref{higgs-energy} and to be accepted as a universal 
initial condition in the JF (thus all the terms in
\eqref{JF-conditions} are of order $\Lambda^4$). 
Then $R^{JF} \sim \Lambda^4\,\Omega^{-2}\,M_{\text{Pl}}^{-2}$ supplants
eq.\,\eqref{JF-curvature} and the reliable initial condition in the
EF (all terms there are always of the same order) 
is obtained with $\Omega\sim
\sqrt{\xi}\,\Lambda\,M_{\text{Pl}}^{-1}$, as follows from \eqref{Higgs-omega}.

To summarize, one concludes that for $R^2$- and Higgs-inflations
formulated in the JF all species of the initial inflaton energy
density should be of the same order. Then the counterparts in the EF
are all of the same order as well.  In this sense these models of
inflation are not as ``unlikely'' as other models with unbounded
potentials.  An improper for the successful inflation initial condition
with a hierarchy between the terms in the EF (e.g. when gradients
dominate) would imply the same hierarchy in the JF, which is unnatural.

\vspace{0.5cm}

We thank F.\,Bezrukov, M.\,Shaposhnikov 
and A.\,Starobinsky for useful correspondence. 
The work was supported by the RSCF grant 14-12-01430.


\end{document}